\documentclass[aps,prl,twocolumn,groupedaddress]{revtex4-1}
\usepackage{graphicx}
\usepackage{amsmath}
\usepackage{amssymb}
\usepackage{braket}

\begin{document}

\title{Steady state thermodynamics in population dynamics}
\author{Yuki Sughiyama, Tetsuya J. Kobayashi}
\affiliation{Institute of Industrial Science, The University of Tokyo, 4-6-1, Komaba, Meguro-ku, Tokyo 153-8505 Japan}
\date{\today}
\begin{abstract}
We report that population dynamics in fluctuating environment accompanies mathematically equivalent structure to steady state thermodynamics.
By employing the structure, population growth in fluctuating environment is decomposed into housekeeping and excess parts. 
The housekeeping part represents the integral of stationary growth rate for each condition during a history of the environmental change. 
The excess part accounts for the excess growth generated when environment is switched. 
Focusing on the excess growth, we obtain Clausius inequality, which gives the upper bound of the excess growth. 
The equality is shown to be achieved in quasistatic environmental changes. 
We also clarify that this bound can be evaluated by "lineage fitness" that is an experimentally observable quantity. 
\end{abstract}
\maketitle
{\it Introduction}- 
Steady state thermodynamics (SST) was established for understanding a ``thermodynamics" of transitions between nonequilibrium steady states (NESS) \cite{01,02,03,04,05,06,07,08,09,10}. 
The core of this theory was proposed by Oono and Paniconi \cite{01} in phenomenological sense, and that is a decomposition of total heat during transitions into housekeeping and excess parts. Former represents the heat dissipated for maintaining NESS, and latter is the heat generated due to relaxation to NESS. 
Based on this decomposition, Clausius inequality is reformulated in nonequilibrium situations. 
While formulated in physics, a mathematically similar framework can be found in biological systems, especially in population dynamics \cite{11,12,13,14,15,16,17,18}. 
This decomposition also contributes evaluation of population growth in fluctuating environment. 

The long term expansion rate of population size (population growth) is the major observable in population dynamics which characterizes the competitive power of the population in evolution. 
In a fixed environment, this quantity converges to a stationary growth rate and it can be evaluated by the largest eigenvalue of time-evolution operator of the population dynamics \cite{17}. 
However, fluctuation of environment disturbs this convergence, and the population growth deviates from the simple integral of the stationary growth rates for each environmental state.
This deviation is the major impact of fluctuating environment (fitness seascape \cite{19,20,21}). 
By employing the same mathematical framework as SST \cite{a1}, we can directly evaluate the deviation as the
excess part of the total population growth (excess growth). 
In application, this decomposition can be exploited to design effective external perturbation to suppress growth of pathogenic and cancer cells \cite{22}. 
In physics, this knowledge can facilitate to design a growing system to physically estimate free energy of a given stochastic system \cite{23}. 

In this letter, we deal with a heterogeneous population of organisms whose type (e.g. geno- and pheno- types) stochastically switches over time.  
We show that the total population growth in fluctuating environment can be generally decomposed into house-keeping and excess parts. 
The house-keeping growth is the integral of the stationary growth rate for each environmental state, and therefore, the excess part accounts for the deviation from the simple integral by dynamic change in the environment. 
If the types of individuals switches by following the detailed balance condition (DBC), the excess growth is shown to satisfy a Clausius inequality in which the entropy function is defined by the stationary measure of the types switching and ``lineage fitness", which implies prosperity of each type in future. 
Clausius {\it equality} is proved for quasistatic cases, and therefore excess growth can be exactly calculated from lineage fitnesses at boundaries of a history of an environmental change.  
In addition, in SST framework, a loss of population growth from the upper bound is translated as ``entropy production", which is evaluated by employing fluctuation relation. 
These results clarify the underlying constraints and thermodynamic structure of the excess growth, and thereby pave the way to further understanding and controlling of the population growth (fitness) under fluctuating environment. 

{\it Setup and Clausius inequality}- 
We consider a simple but general population dynamics that consists of two steps, type switching and duplication. 
Let $x$ be a type of individuals in population, and its switching dynamics is given by a continuous-time ergodic Markov jump process generated by a transition rate $T\left(x|x^{\prime}\right)$, which denotes the jump rate from $x^{\prime}$ to $x$. 
The duplication rate of individuals with type $x$ in environmental condition $y$ is denoted by $\mu_{y}\left(x\right)$. 
From above two steps, the time evolution of the population is described by 
\begin{equation}
\displaystyle \frac{\partial N\left(x,t\right)}{\partial t}=\sum_{x^{\prime}}\left\{\mu_{y_{t}}\left(x\right)\delta_{x,x^{\prime}}+T\left(x|x^{\prime}\right)\right\}N\left(x^{\prime},t\right),\label{PD}
\end{equation}
where $N\left(x,t\right)$ denotes the number of individuals with type $x$. 

Under this setup, we consider population growth during time interval $\left[0,\tau\right]$, which is defined as $\Psi\left[Y\right]\equiv\log\left\{N_{\tau}^{\mathrm{t}\mathrm{o}\mathrm{t}}\left[Y\right]/N_{0}^{\mathrm{t}\mathrm{o}\mathrm{t}}\right\}$. 
Here, $Y$ denotes a history of the environmental change, $Y=\left\{y_{\tau},y_{\tau-\Delta t},...,y_{\Delta t},y_{0}\right\}$ and $N_{t}^{\mathrm{t}\mathrm{o}\mathrm{t}}$ represents the total number of individuals in the population at time $t$, which is evaluated as $N_{t}^{\mathrm{t}\mathrm{o}\mathrm{t}}=\Sigma_{x}N\left(x,t\right)$. 
According to the SST framework, we decompose the population growth as $\Psi\left[Y\right]=\Psi^{\mathrm{h}\mathrm{k}}\left[Y\right]+\Psi^{\mathrm{e}\mathrm{x}}\left[Y\right]$, where $\Psi^{\mathrm{h}\mathrm{k}}$ and $\Psi^{\mathrm{e}\mathrm{x}}$ denote the housekeeping and the excess growth, respectively. 
Here, the housekeeping growth is defined as the integral of the stationary growth rate for each condition during a history $Y$, that is 
\begin{equation}
\displaystyle \Psi^{\mathrm{h}\mathrm{k}}\left[Y\right]\equiv\int_{0}^{\tau}\lambda_{0}\left(y_{t}\right)dt,\label{hkg}
\end{equation}
where $\lambda_{0}\left(y\right)$ represents the stationary growth rate in environmental condition $y$. 
By using Eq. (\ref{PD}), this growth is calculated by the largest eigenvalue of the time-evolution matrix $H_{y}\left(\cdot|\cdot^{\prime}\right)\equiv\mu_{y}\left(\cdot\right)\delta_{\cdot,\cdot^{\prime}}+T\left(\cdot|\cdot^{\prime}\right)$, that is $\Sigma_{x^{\prime}}H_{y}\left(x|x^{\prime}\right)v_{y}\left(x^{\prime}\right)=\lambda_{0}\left(y\right)v_{y}\left(x\right)$ ($\lambda_{0}\left(y\right)>\lambda_{i}\left(y\right)$) \cite{17}. 
Here, $v_{y}\left(x\right)$ is normalized as $\Sigma_{x}v_{y}\left(x\right)=1$ and it represents the largest right eigenvector, which expresses the occupation measure of type $x$ in the stationary state of environmental condition $y$. 
On the other hand, the excess growth is defined by 
\begin{equation}
\displaystyle \Psi^{\mathrm{e}\mathrm{x}}\left[Y\right]\equiv\Psi\left[Y\right]-\int_{0}^{\tau}\lambda_{0}\left(y_{t}\right)dt,
\end{equation}
which implies the deviation from the integral of stationary growth rate. 
That is, this growth represents the growth generated when environment is switched. 
In this study, we assume that the transition rate $T\left(\cdot|\cdot^{\prime}\right)$ satisfies the detailed balance condition (DBC) \cite{24}, $T\left(x|x^{\prime}\right)P_{T}^{\mathrm{s}\mathrm{t}}\left(x^{\prime}\right)=T\left(x^{\prime}|x\right)P_{T}^{\mathrm{s}\mathrm{t}}\left(x\right)$, where $P_{T}^{\mathrm{s}\mathrm{t}}$ denotes the stationary measure of $T\left(\cdot|\cdot^{\prime}\right)$. 
This assumption is not so restrictive biologically since geno- and pheno-type switching dynamics are often described by using this condition \cite{15,25}. 
(We also discuss non-DBC cases in {\it Discussion} \cite{26}.) 
As shown in the following section, by employing the DBC, we obtain Clausius inequality, 
\begin{equation}
\Psi^{\mathrm{e}\mathrm{x}}\left[Y\right]\leq S\left(y_{\tau}\right)-S\left(y_{0}\right),\label{CIE}
\end{equation}
where $S\left(y\right)$ represents ``entropy" in population dynamics \cite{27}, which is given by
\begin{equation}
S\displaystyle \left(y\right)=\frac{1}{2}\log\sum_{x}P_{T}^{\mathrm{s}\mathrm{t}}\left(x\right)u_{y}\left(x\right).\label{ent}
\end{equation}
Here, $u_{y}\left(x\right)$ denotes the largest left eigenvector, $\Sigma_{x}u_{y}\left(x\right)H_{y}\left(x|x^{\prime}\right)=u_{y}\left(x^{\prime}\right)\lambda_{0}\left(y\right)$, with normalization condition $\Sigma_{x}u_{y}\left(x\right)v_{y}\left(x\right)=1$; and $P_{T}^{\mathrm{s}\mathrm{t}}$ represents the stationary measure of the type-switching rate $T\left(\cdot|\cdot^{\prime}\right)$. 
$u_{y}\left(x\right)$ also represents ``lineage fitness" of type $x$ in environmental condition $y$, which implies prosperity of each type in future \cite{13,14,15,16}. 
In addition, we can prove that $S\left(y\right)\leq 0$ for arbitrary $y$ and $S\left(y\right)=0$ iff $y$ is no selection situation, i.e. $\mu_{y}\left(x\right)=\mathrm{const}.$ for arbitrary $x$.
(Proof is shown in Supplement C). 
The equality of Eq. (\ref{CIE}) is achieved in quasistatic environmental changes. 
This entropy can be interpreted and experimentally observed as follows. 

Suppose that the population is in the stationary growing situation with a fixed environmental condition $y$, that is, the population has the stationary occupation measure $v_{y}$. 
If we track the offsprings of individuals with type $x$ at a initial time, $t=0$, the offsprings will change their types and grow in the population. 
The fraction of the offsprings (irrespective of their types) in the population changes over time and finally converges to some value at $ t\rightarrow\infty$ (say $P_{R_{y}}^{\mathrm{s}\mathrm{t}}\left(x\right)$). 
Then, the lineage fitness of type $x$ is given as $u_{y}\left(x\right)=P_{R_{y}}^{\mathrm{s}\mathrm{t}}\left(x\right)/v_{y}\left(x\right)$, see FIG. 1. 
Accordingly,  if we employ a labeling technique with which individuals with specific types can be labeled and their offsprings inherit the label, we can estimate the lineage fitness experimentally by measuring the fraction of the labeled offsprings in the population. 
Furthermore, it is known that the convergence fraction $P_{R_{y}}^{\mathrm{s}\mathrm{t}}\left(x\right)$ \cite{a2} is given by the stationary measure of the retrospective processes $R_{y}$ \cite{13,14,15,16,17,18} (see Eq. (5) in Supplement A). 
Thus, we can calculate the lineage fitness by tracing the linage of the growing population time-backwardly. 
Next, we consider how to observe $P_{T}^{\mathrm{s}\mathrm{t}}\left(x\right)$ in experiments. 
Since $P_{T}^{\mathrm{s}\mathrm{t}}$ is the stationary measure of the type-switching process $T$, we can obtain it by tracing the linage of the population time-forwardly (see Supplement A). 
Taking these facts into account, we find that the entropy (\ref{ent}) can be evaluated in experiments. 
\begin{figure}[h]
\begin{center}
\includegraphics*[height=4cm]{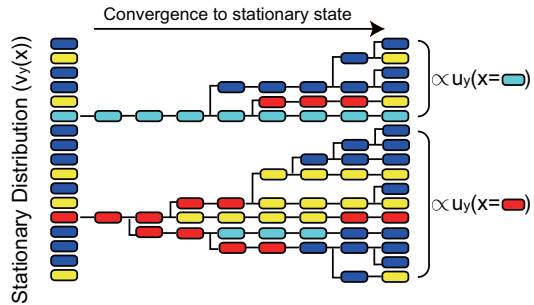}
\caption{(color online). 
Types of individuals are represented by colors. 
In experiments, we can calculate the lineage fitness by the following steps; 
(i) We observe the fraction of individuals with type $x$ at initial time, that is the occupation measure $v_{y}\left(x\right)$; 
(ii) We cultivate the population in a fixed environmental condition $y$ for a sufficiently long time; 
(iii) We finally observe the fraction of the offspring of the individuals with type $x$ at initial time, which is denoted as $P_{R_{y}}^{\mathrm{s}\mathrm{t}}\left(x\right)$; 
(iv) We obtain the lineage fitness of type $x,\ u_{y}\left(x\right)$, by the ratio, $u_{y}\left(x\right)=P_{R_{y}}^{\mathrm{s}\mathrm{t}}\left(x\right)/v_{y}\left(x\right)$. }
\end{center}
\end{figure}

Before working on the derivation of Clausius inequality (\ref{CIE}), we consider the quasi-static situation and ``entropy production". 
In quasistatic environmental changes,Clausius {\it equality}, $\Psi^{\mathrm{e}\mathrm{x}}\left[Y\right]=S\left(y_{\tau}\right)-S\left(y_{0}\right)$, is achieved. 
Therefore, the excess growth can be evaluated by observing entropies at boundaries of an environment-switching history $Y$, although it is functional of the history $Y$. 
On the other hand, in non-quasistatic situations, ``entropy production" occurs, and it is defined as $\sigma\left[Y\right]\equiv\left\{S\left(y_{\tau}\right)-S\left(y_{0}\right)\right\}-\Psi^{\mathrm{e}\mathrm{x}}\left[Y\right]$. 
In terms of population dynamics, this implies a loss of population growth from the upper bound in non-quasistatic cases. 
As shown in the following section, this entropy production is evaluated by Kawai-Parrondo-Van den Broeck type fluctuation theorem \cite{28}, 
\begin{equation}
\displaystyle \sigma\left[Y\right]=D^{\mathrm{s}\mathrm{y}\mathrm{m}}\left[Y\right]+\sum_{x}v_{y_{\tau}}\left(x\right)\log u_{y_{\tau}}\left(x\right),\label{ep}
\end{equation}
where $D^{\mathrm{s}\mathrm{y}\mathrm{m}}$ represents the symmetrized Kullback-Leibler divergence, 
\begin{equation}
D^{\mathrm{s}\mathrm{y}\mathrm{m}}\displaystyle \left[Y\right]\equiv\sum_{X}P_{B}\left[X|Y\right]\log\frac{P_{B}\left[X|Y\right]}{P_{R}^{1/2}\left[X|Y\right]P_{\tilde{R}}^{1/2}\left[\tilde{X}|\tilde{Y}\right]}.\label{Dsym}
\end{equation}
The definition and a meaning of path probabilities, $P_{B},\ P_{R}$ and $P_{\tilde{R}}$, are shown in Supplement A. 

{\it Derivation of Clausius inequality}- 
Let us begin with derivation of the {\it equality}, i.e. we consider quasistatic environmental change. 
For simplicity, we rewrite Eq. (\ref{PD}) by using bra-ket notation as
\begin{equation}
\displaystyle \frac{\partial}{\partial t}\ket{N\left(t\right)}=\hat{H}_{y_{t}}\ket{N\left(t\right)},
\end{equation}
where $\hat{H}_{y}$ represents the time-evolution operator, $\hat{H}_{y}=\hat{\mu}_{y}+\hat{T}$ and $\braket{x|\hat{H}_{y}|x^{\prime}} =H_{y}\left(x|x^{\prime}\right)$; 
$\ket{N\left(t\right)}$ denotes the population vector, that is $\braket{x|N\left(t\right)} =N\left(x,t\right)$. 
In this notation, the population growth $\psi$ is expressed by 
\begin{equation}
e^{\Psi\left[Y\right]}=\Braket{\mathcal{P}|T\exp\left[\int_{0}^{\tau}\hat{H}_{y_{t}} dt\right]|N\left(0\right)},\label{d1}
\end{equation}
where $T\exp\left[\cdot\right]$ denotes the time-ordered exponential and $\bra{\mathcal{P}}$ is defined as $\braket{\mathcal{P}|x} =1$ for any $\ket{x}$. 
We write eigenvalues of $\hat{H}_{y}$ as $\lambda_{i}\left(y\right)$, where $i=0$ indicate the largest eigenvalue. 
In addition, we prepare left and right eigenvectors as $\bra{\lambda_{i}\left(y\right)}$ and $\ket{\lambda_{i}\left(y\right)}$, respectively. 
These vectors are normalized as $\braket{\lambda_{i}\left(y\right)|\lambda_{j}\left(y\right)} =\delta_{ij}$ and $\braket{\mathcal{P}|\lambda_{0}\left(y\right)} =1$. 
Thus, we can write the stationary occupation measure $v_{y}$ and lineage fitness $u_{y}$ as $v_{y}\left(x\right)= \braket{x|\lambda_{0}\left(y\right)}$ and $u_{y}\left(x\right)= \braket{\lambda_{0}\left(y\right)|x}$, respectively. 
By substituting the completeness relation for eigenvectors $\Sigma_{i} \ket{\lambda_{i}\left(y\right)}\bra{\lambda_{i}\left(y\right)} =1$ into Eq. (\ref{d1}), we have 
\begin{eqnarray}
\displaystyle \nonumber&& e^{\Psi\left[Y\right]}=\sum_{i_{\tau},i_{\tau-\Delta t},...,i_{0}}\braket{\mathcal{P}|\lambda_{i_{\tau}}\left(y_{\tau}\right)}\\
\nonumber&&\times\braket{\lambda_{i_{\tau}}\left(y_{\tau}\right)|e^{\hat{H}_{y_{\tau}}\Delta t}|\lambda_{i_{\tau-\Delta t}}\left(y_{\tau-\Delta t}\right)}\times\cdots\\
&&\times\braket{\lambda_{i_{\Delta t}}\left(y_{\Delta t}\right)|e^{\hat{H}_{y_{\Delta t}}\Delta t}|\lambda_{i_{0}}\left(y_{0}\right)}\braket{\lambda_{i_{0}}\left(y_{0}\right)|N\left(0\right)}.
\end{eqnarray}
By assuming that initial population is stationary, that is $\ket{N_{0}} = \ket{\lambda_{0}\left(y_{0}\right)}$, and by taking into account that environmental change is quasistatic, we reach 
\begin{eqnarray}
\nonumber e^{\Psi\left[Y\right]}&=&\braket{\lambda_{0}\left(y_{\tau}\right)|e^{\hat{H}_{y_{\tau}}\Delta t}|\lambda_{0}\left(y_{\tau-\Delta t}\right)}\times\cdots\\
&&\times\braket{\lambda_{0}\left(y_{\Delta t}\right)|e^{\hat{H}_{y_{\Delta t}}\Delta t}|\lambda_{0}\left(y_{0}\right)},\label{d2}
\end{eqnarray}
where we employ the adiabatic approximation. 
By taking logarithm to both sides of Eq. (\ref{d2}), we obtain the population growth, 
\begin{equation}
\displaystyle \Psi\left[Y\right]=\int_{0}^{\tau}\lambda_{0}\left(y_{t}\right)dt-\int_{0}^{t}dt\braket{\lambda_{0}\left(y\right)|\nabla_{y}|\lambda_{0}\left(y\right)}\cdot\dot{y},\label{d3}
\end{equation}
where we use $\braket{\lambda_{0}\left(y_{t+\Delta t}\right)|\lambda_{0}\left(y_{t}\right)} =e^{-\braket{\lambda_{0}\left(y\right)|\nabla_{y}|\lambda_{0}\left(y\right)}\cdot\dot{y}\Delta t}$ and $\nabla_{y}$ denotes differentiation with respect to $y$. 
The dots $\cdot$ and $\dot{y}$ represent inner product and time differentiation, and thus $\dot{y}\cdot\nabla_{y}=\Sigma_{i}\left(dy_{i}/dt\right)\left(\partial/\partial y_{i}\right)$ where $i$ expresses the dimension of environment. 
Taking into account the definition of the housekeeping growth, Eq. (\ref{hkg}), we can evaluate the excess growth by the second term of Eq. (\ref{d3}), $\Psi^{\mathrm{e}\mathrm{x}}\left[Y\right]=- \int_{0}^{t} dt \braket{\lambda_{0}\left(y\right)|\nabla_{y}|\lambda_{0}\left(y\right)} \cdot\dot{y}$. 
This representation implies that the excess growth can be given by the geometric phase (Berry phase) \cite{21,29,30,31}. 
By using the completeness relation $\Sigma_{x} \ket{x}\bra{x} =1$, we find a more familiar form without bra-ket notation, 
\begin{equation}
\displaystyle \Psi^{\mathrm{e}\mathrm{x}}\left[Y\right]=-\int_{0}^{t}dt\dot{y}\cdot\sum_{x}u_{y}\left(x\right)\nabla_{y}v_{y}\left(x\right).\label{d4}
\end{equation}
Next, by using the DBC assumption for the type-switching operator $\hat{T}$, we calculate the integrand in Eq. (\ref{d4}) by a potential condition. 
From DBC, we can obtain a relation between the stationary occupation measure $v_{y}$ and the lineage fitness $u_{y}$ as 
\begin{equation}
C\left(y\right)v_{y}\left(x\right)=P_{T}^{\mathrm{s}\mathrm{t}}\left(x\right)u_{y}\left(x\right),\label{d5}
\end{equation}
where the constant $C\left(y\right)$ is given by $C\left(y\right)=\Sigma_{x}P_{T}^{\mathrm{s}\mathrm{t}}\left(x\right)u_{y}\left(x\right)=\Sigma_{x}P_{T}^{\mathrm{s}\mathrm{t}}\left(x\right)u_{y}^{2}\left(x\right)$. 
(Derivation of Eq. (\ref{d5}) is shown in Supplement B.) 
By using these normalization and Eq. (\ref{d5}), we have 
\begin{equation}
\displaystyle \nabla_{y}\left\{\frac{1}{2}\log C^{-1}\left(y\right)\right\}=\sum_{x}u_{y}\left(x\right)\nabla_{y}v_{y}\left(x\right).\label{d6}
\end{equation}
By substituting Eq. (\ref{d6}) into Eq. (\ref{d4}), we obtain Clausius {\it equality}, 
\begin{equation}
\Psi^{\mathrm{e}\mathrm{x}}\left[Y\right]=S\left(y_{\tau}\right)-S\left(y_{0}\right),\label{CE}
\end{equation}
where $S\left(y\right)$ denotes ``entropy" defined in Eq. (\ref{ent}), because $S\left(y\right)=\left(1/2\right)\log C\left(y\right)$. 

In the following part, we derive Clausius {\it inequality}, i.e. we prove that Eq. (\ref{CE}) gives the upper bound of the excess growth. 
Consider a transition between stationary growing situations at environmental condition $y_{0}$ and $y_{\tau}$. 
We also assume that a history of environmental change, $Y,$ is non-quasistatic during this transition. 
By employing two kinds of detailed fluctuation relation (FR) constructed in the population dynamics Eq. (\ref{PD}) and DBC assumption for type switching $T$, we obtain 
\begin{eqnarray}
\displaystyle \nonumber&&\Psi^{\mathrm{e}\mathrm{x}}\left[Y\right]+D^{\mathrm{s}\mathrm{y}\mathrm{m}}\left[Y\right]+\sum_{x}v_{y_{\tau}}\left(x\right)\log u_{y_{\tau}}\left(x\right)\\
&&=S\left(y_{\tau}\right)-S\left(y_{0}\right),\label{add1}
\end{eqnarray}
where $D^{\mathrm{s}\mathrm{y}\mathrm{m}}\left[Y\right]$ is the symmetrized Kullback-Leibler divergence defined in Eq. (\ref{Dsym}). 
(Derivation of FRs and Eq. (\ref{add1}) is shown in Supplement D.) 
Therefore, if $D^{\mathrm{s}\mathrm{y}\mathrm{m}}\left[Y\right]+\Sigma_{x}v_{y_{\tau}}\left(x\right)\log u_{y_{\tau}}\left(x\right)\geq 0$, we can obtain Clausius inequality (\ref{CIE}). 
To prove the above inequality, we suppose that the final environmental condition $y_{\tau}=y_{F}$ makes no selection situation, i.e. $\mu_{y_{F}}\left(x\right)=\mathrm{const}.$ and thus $u_{y_{F}}\left(x\right)=1$. 
By taking arbitrary intermediate environmental condition $y_{M}$, we consider maximum excess growths within intervals, $y_{0}$ to $y_{M},\ y_{M}$ to $y_{F}$, and $y_{0}$ to $y_{F}$, also see FIG. 2; 
We denotes these maximum growths as  $\Psi^{\mathrm{e}\mathrm{x}}\left[Y_{0\rightarrow M}^{*}\right],\ \Psi^{\mathrm{e}\mathrm{x}}\left[Y_{M\rightarrow F}^{*}\right]$ and $\Psi^{\mathrm{e}\mathrm{x}}\left[Y_{0\rightarrow F}^{*}\right]$, where $Y_{i\rightarrow j}^{*}$ represents the history which maximizes the excess growth functional $\Psi^{\mathrm{e}\mathrm{x}}\left[\cdot\right]$ in an interval $\left[y_{i},y_{j}\right]$. 
By using  $\Sigma_{x}v_{y_{F}}\left(x\right)\log u_{y_{F}}\left(x\right)=0$ and $S\left(y_{F}\right)=0$, we can evaluate these growths as $\Psi^{\mathrm{e}\mathrm{x}}\left[Y_{0\rightarrow M}^{*}\right]=\left\{S\left(y_{M}\right)-S\left(y_{0}\right)\right\}-\left\{D^{\mathrm{s}\mathrm{y}\mathrm{m}}\left[Y_{0\rightarrow M}^{*}\right]+\Sigma_{x}v_{y_{M}}\left(x\right)\log u_{y_{M}}\left(x\right)\right\},$ 
$\Psi^{\mathrm{e}\mathrm{x}}\left[Y_{M\rightarrow F}^{*}\right]=-S\left(y_{M}\right)$ and $\Psi^{\mathrm{e}\mathrm{x}}\left[Y_{0\rightarrow F}^{*}\right]=-S\left(y_{M}\right)$, where we use Eq. (\ref{add1}). 
Since $\Psi^{\mathrm{e}\mathrm{x}}\left[Y_{0\rightarrow M}^{*}\right]+\Psi^{\mathrm{e}\mathrm{x}}\left[Y_{M\rightarrow F}^{*}\right]\leq\Psi^{\mathrm{e}\mathrm{x}}\left[Y_{0\rightarrow F}^{*}\right]$ is satisfied, we obtain $D^{\mathrm{s}\mathrm{y}\mathrm{m}}\left[Y_{0\rightarrow M}^{*}\right]+\Sigma_{x}v_{y_{M}}\left(x\right)\log u_{y_{M}}\left(x\right)\geq 0$. 
Then, since $Y_{0\rightarrow M}^{*}$ gives the maximum of $\Psi^{\mathrm{e}\mathrm{x}}\left[\cdot\right]$, it also attains the minimum of the functional $D^{\mathrm{s}\mathrm{y}\mathrm{m}}\left[\cdot\right]+\Sigma_{x}v_{y_{M}}\left(x\right)\log u_{y_{M}}\left(x\right)$. 
Accordingly, for an arbitrary history $Y,\ D^{\mathrm{s}\mathrm{y}\mathrm{m}}\left[Y\right]+\Sigma_{x}v_{y_{\tau}}\left(x\right)\log u_{y_{\tau}}\left(x\right)\geq 0$ is satisfied. 
As a result, from Eq. (\ref{add1}), we find Clausius inequality, 
\begin{equation}
\Psi^{\mathrm{e}\mathrm{x}}\left[Y\right]=\sigma\left[Y\right]+\left\{S\left(y_{\tau}\right)-S\left(y_{0}\right)\right\}\leq S\left(y_{\tau}\right)-S\left(y_{0}\right),
\end{equation}
where $\sigma\left[Y\right]$ denotes the entropy production defined in Eq. (\ref{ep}). 
\begin{figure}[h]
\begin{center}
\includegraphics*[height=2.5cm]{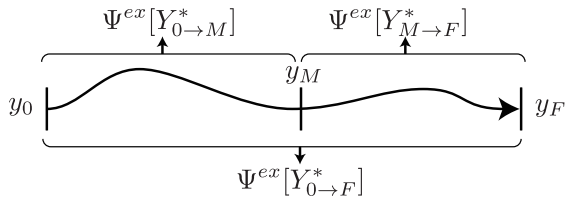}
\caption{Three maximum excess growths.}
\end{center}
\end{figure}

{\it Discussion}- 
We have established SST structure in population dynamics. 
Owing to Clausius inequality, the upper bound of excess growth is evaluated by the lineage fitnesses of initial and final environmental conditions. 
However, we must recall that the DBC is assumed in our theory. 
When we deal with non-DBC type switching (e.g. metabolic switching and circadian rhythm), Clausius inequality no longer available. 
Even for quasistatic environmental change, we need directly to calculate geometric phase (\ref{d4}) in the similar way as T\u{a}nase-Nicolafs study \cite{21}, because excess growth can not be evaluated by the potential as in Clausius equality (\ref{CE}). 
Furthermore, in non-DBC situations, it is still uncertain whether geometric phase (\ref{d4}) gives upper bound of the excess growth or not \cite{a3}. 
These are open problems in this study. 

{\it Acknowledgments}-
This research is partially supported by the JST PRESTO program; Platform for Dynamic Approaches to Living System from the Ministry of Education, Culture, Sports, Science and Technology, Japan; and the Specially Promoted Project of the Toyota Physical and Chemical Research Institute. 



\begin{thebibliography}{00}
\bibitem{a1}  We focus only on mathematical similarity between mathematics of SST and that of population dynamics.
\bibitem{26}  Physicists may claim that SST framework is not needed for systems with the DBC. 
However, we now deal with population dynamics differently from stochastic processes without population growth which are often employed in nonequilibrium physics.
Thus, SST framework plays an essential role even for the DBC.
\bibitem{27}  By taking into account the form of Eq. (\ref{ent}), we should refer to this quantity as ``free energy". However, to clarify correspondence between our theory and SST, we call it ``entropy" in this letter.
\bibitem{a2} $P_{R_{y}}^{\mathrm{s}\mathrm{t}}$ is called ancestral distribution in Ref. \cite{13,14,15,16,17,18}.
\end{thebibliography}
\end{document}